\documentclass[10pt,twocolumn,letterpaper]{article}

\usepackage[final]{cvpr}      
\usepackage{multirow}
\usepackage{multicol}
\usepackage{setspace}
\usepackage[accsupp]{axessibility}

\usepackage[dvipsnames]{xcolor}

\definecolor{cvprblue}{rgb}{0.21,0.49,0.74}
\usepackage[pagebackref,breaklinks,colorlinks,citecolor=cvprblue]{hyperref}
\usepackage{colortbl,xcolor,array}

\def\paperID{*} 
\def\confName{CVPR}
\def\confYear{2024}
\def\authorBlock{
	Xin Wang$^{1} $ \qquad Lizhi Wang$^{1}$\thanks{Corresponding Author: Lizhi Wang (wanglizhi@bit.edu.cn)} \qquad Xiangtian Ma$^{1}$ \qquad Maoqing Zhang$^{1}$ \qquad Lin Zhu$^{1}$ \qquad Hua Huang$^{2}$ \\
	$^{1}$Beijing Institute of Technology \qquad $^{2}$	Beijing Normal University \\}

\newif\ifreview 
\newif\ifarxiv 
\newif\ifcamera 
\newif\ifrebuttal

\title{In2SET: Intra-Inter Similarity Exploiting Transformer for  \\ Dual-Camera Compressive Hyperspectral Imaging}

\begin{document}
\author{\authorBlock}
\maketitle
\begin{abstract}
	Dual-camera compressive hyperspectral imaging (DCCHI) offers the capability to reconstruct 3D hyperspectral image (HSI) by fusing compressive and panchromatic (PAN) image,  which has shown great potential for snapshot hyperspectral imaging in practice. In this paper, we introduce a novel DCCHI reconstruction network, intra-inter similarity exploiting Transformer (In2SET). Our key insight is to make full use of the PAN image to assist the reconstruction. To this end, we propose to use the intra-similarity within the PAN image as a proxy for approximating the intra-similarity in the original HSI, thereby offering an enhanced content prior for more accurate HSI reconstruction. Furthermore, we propose to use the inter-similarity to align the features between HSI and PAN images, thereby maintaining semantic consistency between the two modalities during the reconstruction process. By integrating In2SET into a PAN-guided deep unrolling (PGDU) framework, our method substantially enhances the spatial-spectral fidelity and detail of the reconstructed images, providing a more comprehensive and accurate depiction of the scene. Experiments conducted on both real and simulated datasets demonstrate that our approach consistently outperforms existing state-of-the-art methods in terms of reconstruction quality and computational complexity. The code is available at  \url{https://github.com/2JONAS/In2SET}.
\end{abstract}

\section{Introduction}
\label{sec:intro}

\begin{figure}[t!]
	\centering
	\includegraphics[width=3.20in,height=1.60in]{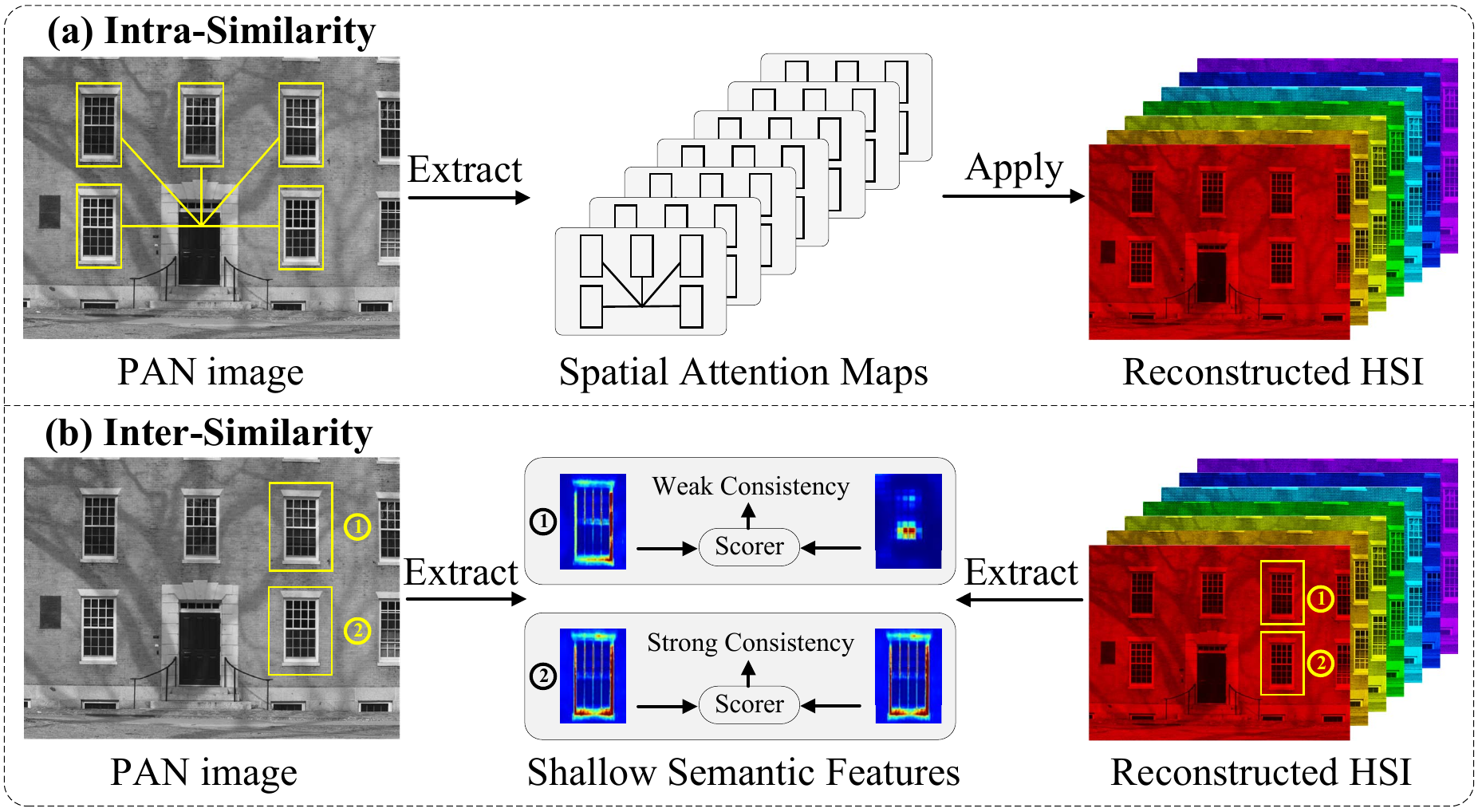}
	\caption{Illustration of the proposed In2SET method for hyperspectral image reconstruction. (a) Intra-Similarity: extraction and application of spatial attention maps from the PAN image to enhance the spatial resolution of the reconstructed HSI. (b) Inter-Similarity: utilization of semantic features from the HSI and PAN image, scored by their consistency, to inform and refine the reconstruction of HSI.} 
	\label{fig:idea}
\end{figure}

Hyperspectral imaging collects and processes scene information by dividing the whole spectrum into tens or hundreds of bands~\cite{zh23ha,li14sp,do19de}. Owing to the capability of detailed scene representation, this technique has been widely adopted in many fields, e.g., medical diagnosis, health care, remote sensing~\cite{li03re,ca14hy,bi13hy,sh19hy}, and various computer vision tasks, e.g., recognition, classification, segmentation~\cite{pa03fa,xu20at,li11hy}.

The hyperspectral image (HSI) inherently possesses a three-dimensional (3D) structure, comprising two dimensions of spatial information and one dimension of spectral information. Consequently, existing 2D imaging sensors lack the capability to directly capture this complete 3D signal. With the rapid progress in computational photography, snapshot spectral imagers have been developed to enable hyperspectral imaging in dynamic scenes~\cite{yu21sn,la15co,ca16co}. The coded aperture snapshot spectral imaging (CASSI)~\cite{ge07si,wa08si,ar13co}, serving as a representative prototype of snapshot spectral imagers, captures one compressive image that encodes the spectral information by leveraging a decorated optical system. Inspired by multi-modal image fusion techniques, dual-camera compressive hyperspectral imaging (DCCHI)~\cite{wa15du,wa15hi,wa18hi} upgrades the original CASSI by adding a panchromatic (PAN) camera to supply spatial information lost in CASSI. As a result, DCCHI can achieve more accurate reconstructions of HSI by fusing the compressive and PAN images compared to CASSI, which has shown great potential for snapshot hyperspectral imaging in practice.

The significant challenge of DCCHI lies in the HSI reconstruction method, where the key factor that determines the reconstruction performance is the ability to model HSI priors. Early methods mainly rely on general image priors such as smoothness\cite{bi07an,yu16ge,wa15du}, sparsity\cite{fi07gr,yu15co,wa16si}, low-rank\cite{fu16ex,li18ra,zhsp19co}, and non-local similarity\cite{wa17ad,he19no}, which ignore the continuous and complete spatial characteristics of PAN images, a feature that could provide stronger image priors.

In recent years, some PAN-guided methods like ANSR~\cite{wa19hy,wa17ad}, PFusion~\cite{he21fa}, and PIDS~\cite{ch23pr} have utilized PAN image as prior information to guide the reconstruction of HSI. While these approaches have demonstrated merit, their performance is limited by the following two reasons: firstly, current PAN-guided methods rely on handcrafted prior constraints, which lack a precise depiction of the intrinsic structures of HSI and often require manual parameter tuning. Consequently, when dealing with complex scenes, these methods tend to produce degraded reconstruction quality. Secondly, current PAN-guided methods typically model inter-modality correlations between PAN and HSI images from a single perspective, such as using edge information from the PAN images to constrain the HSI edges~\cite{ch23pr} or employing PAN image content to enrich HSI spatial detail~\cite{he21fa}. However, HSI exhibits multifaceted consistency with PAN images across various aspects such as structure, texture, luminance, contrast, and semantic content. Relying on one constraint may not fully exploit the comprehensive spatial-semantic guidance offered by the PAN image. In summary, accurately leveraging PAN image-guided prior modeling for HSI reconstruction is the key to further improving reconstruction quality.

In this paper, we propose an intra-inter similarity exploiting Transformer (In2SET) for DCCHI. The proposed In2SET module is integrated into a PAN-guided deep unrolling (PGDU) framework for reconstruction. The PGDU makes use of a multi-scale feature pyramid extracted from the PAN image as guidance. This feature pyramid provides hierarchical representations of the PAN at different resolutions, allowing the network to capture both global context and fine details, resulting in improved spatial fidelity and detail in the reconstructed images.

Our key insight is to make full use of the PAN image to assist the reconstruction. The In2SET is based on two observations: (1) given the large redundancy in both the spatial and the spectral dimensions, intra-similarity exist ubiquitously within HSI. Since the original HSI is not available in practice, intra-similarity needs to be computed based on intermediate reconstruction results. However, relying on these intermediate results for intra-similarity calculations introduces unreliability. Fortunately, the PAN image can be regarded as the integral of all spectral bands and provides a natural reference for estimating the intra-similarity in the original HSI. Thus, we propose to employ the intra-similarity in the PAN image as an approximation proxy for the intra-similarity in the original HSI, which provides an enhanced content prior for HSI reconstruction. (2) Since the HSI and the PAN image describe the same scenes, inter-similarity should exist between the HSI and the PAN image, considering semantic information. Inter-similarity, serving as a scorer, assigns higher weights to features in areas of the HSI that are more similar to the corresponding areas in the PAN image, thereby improving feature representation in those areas and minimizing uncertainty and risk in the reconstruction process. 
This capability enables the reconstruction of the highly ill-posed inverse problem in a more confidently supervised manner, providing new contextual information for HSI reconstruction.

In a nutshell, this work integrates intra-similarity and inter-similarity with the advanced Transformer mechanism, specifically for DCCHI, resulting in high-fidelity HSI reconstruction. The main contributions of this work are as follows:

\begin{itemize}[leftmargin=2em]
	\item We propose the PGDU, which employs the feature pyramid from the PAN image to guide the reconstruction of the HSI.
	\item We propose the In2SET, which employs a novel attention mechanism, specifically designed to concurrently capture both intra-similarity and inter-similarity between spectral and PAN images.
	\item Experiments on real and simulated data demonstrate that our method consistently outperforms state-of-the-art approaches.
\end{itemize}
\section{Related Work}
\label{sec:relate work}
This section provides an overview of the advancements in HSI reconstruction, focusing on three pivotal areas: hand-crafted HSI priors modeling, data-driven HSI priors modeling, and PAN-guided HSI reconstruction.

\subsection{Hand-Crafted HSI Priors Modeling}
\label{sec:hand-crafted image priors}
Early HSI reconstruction relied heavily on hand-crafted priors. Total Variation (TV) regularization~\cite{wa15du,zhmq19co,zh18fa} utilized the piecewise smoothness of images to constrain TV, albeit at the risk of suppressing important high-frequency structural details. Sparse representation techniques~\cite{ar13hi,yu15co,wa16si} have been employed, focusing on constructing dictionaries to model image sparsity. Low-rank priors~\cite{li18ra,zhsp19co,wa21ad} have been applied, designed to capture contextual information in high-dimensional data. However, these hand-crafted methods often required extensive manual parameter tuning and struggled to represent complex data scenarios effectively.

\subsection{Data-Driven HSI Priors Modeling}
\label{sec:Data-driven Methods}
The advent of deep learning has revolutionized the field of HSI reconstruction. Models such as HyperReconNet~\cite{wa18hy}, $\lambda$-net~\cite{mi19la}, TSA-Net~\cite{me20en}, DNU~\cite{wa20dn}, DGSMP~\cite{hu21de} and PnP-DIP-HSI~\cite{me21se} have markedly improved reconstruction quality by modeling HSI data prior knowledge. Furthermore, the introduction of Transformer-based methods like MST~\cite{ca22ma}, CST~\cite{ca22co}, DAU~\cite{ca22de}, RDLUF~\cite{do23re}, and PADUT~\cite{li23pi} has led to even more significant advances in reconstruction performance. Benefiting from the attention mechanism, Transformers can efficiently model both spatial and spectral similarities within HSI data, enabling a more flexible handling of input data and providing a richer feature representation for HSI reconstruction.

\subsection{PAN-Guided HSI Reconstruction}

The integration of PAN guidance in HSI reconstruction represents a significant stride in addressing the inherent limitations of traditional CASSI. ANSR\cite{wa19hy,wa17ad} proposed an adaptive non-local sparse representation model, guided by the PAN image. PFusion\cite{he21fa} leveraged RGB measurements for spatial coefficient estimation and CASSI measurements for spectral basis acquisition. It features a patch processing strategy to enhance the spectral low-rank property, optimizing the model efficiently without iterative methods or a spectral sensing matrix. PIDS\cite{ch23pr} utilized the RGB measurement as a prior image to enhance semantic correspondence between HSI and the PAN image. These methods have marked significant advancements in the field. However, they often require manual parameter tuning and fail to leverage the rich spatial context and detailed features embedded within the PAN image, suggesting potential areas for further improvements in reconstruction.

Our motivation is to utilize the spatial semantic information provided by the PAN image to improve the HSI reconstruction result in DCCHI. We aim to develop a method that combines the strengths of PAN guidance to model the intra-inter similarity for more accurate reconstruction.

\section{PGDU}
\label{sec:DCCHI}
\subsection{DCCHI Forward Model}
\begin{figure}[t!]
	\centering
	\includegraphics[width=3.0in,height=2.25in]{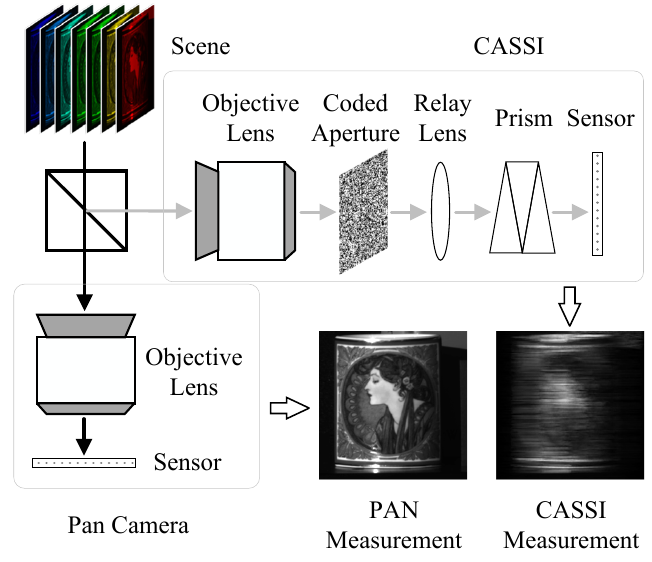} 
	\caption{The dual-camera compressive hyperspectral imaging system.} 
	\label{fig:DCCHI}
        \vspace{-4.5mm}
\end{figure}

DCCHI consists of a beam splitter, a CASSI branch, and a PAN camera branch, as shown in Figure~\ref{fig:DCCHI}. The beam splitter divides the incident light from the scene into two equal-intensity parts. One part of the light is captured by CASSI, which undergoes spatial modulation and spectral modulation to form the CASSI measurement. The other part of the light is directly captured by the PAN camera.

Let \(\mathcal{X} \in \mathbb{R}^{H \times W \times C}\) represent the spectral image of the target scene, where \(H\), \(W\), and \(C\) are the height, width, and number of spectral bands, respectively. The spatial and spectral modulation in CASSI is given by
\begin{equation} 
	\label{eq31}
	\mathbf{Y^{c}} = \sum_{c=1}^{C} \mathrm{shift}(\mathbf{\phi} \odot \mathcal{X}[:,:,c]) + \mathbf{V^c},
\end{equation}
where \(\mathbf{\phi}\) is the transmission intensity of the coded aperture, represented as \(\mathbf{\phi} \in \mathbb{R}^{H \times W}\), \(\mathrm{shift(\cdot)}\) indicates the spectral modulation caused by the dispersion prism, \(\mathbf{Y^{c}}\) is the image captured by the CASSI branch, and \(\mathbf{V^c}\) is regarded as Gaussian noise in CASSI.

To simplify to a matrix-vector form, the imaging model of the CASSI branch is represented as
\begin{equation}
	\label{eq33}
	\mathbf{y^{c}} =  \mathbf{\Phi^{c}} \mathbf{x} + \mathbf{v^c},
\end{equation}
where $\mathbf{\Phi^{c}}$ denotes the imaging model of the CASSI branch sensing matrix, $\mathbf{y^{c}}$ is the vector form of $\mathbf{Y^{c}}$, $\mathbf{x}$ is the vector form of $\mathcal{X}$, and $\mathbf{v^c}$ is the vector form of $\mathbf{V^c}$.

Correspondingly, the imaging model of the PAN camera branch is described by
\begin{equation}
	\label{eq35}
	\mathbf{y^{p}} =  \mathbf{\Phi^{p}} \mathbf{x}  + \mathbf{v^p},
\end{equation}
where \(\mathbf{\Phi^{p}}\) denotes the imaging model of the PAN camera branch sensing matrix, determined by the spectral response of the sensor, $\mathbf{y^{c}}$ is the vector form of PAN measurement, and \(\mathbf{v^p}\) is regarded as Gaussian noise in the PAN measurement. To connect the imaging models to the subsequent discussion, we define \(\mathbf{y}\), \(\boldsymbol{\Phi}\), and \(\mathbf{v}\) as
\begin{equation}
	\label{eq36}
	\mathbf{y} = 
	\begin{bmatrix}
		\mathbf{y^{c}} \\
		\mathbf{y^{p}}
	\end{bmatrix}, \quad
	\boldsymbol{\Phi} = 
	\begin{bmatrix}
		\mathbf{\Phi^{c}} \\
		\mathbf{\Phi^{p}}
	\end{bmatrix}, \quad
	\mathbf{v} = 
	\begin{bmatrix}
		\mathbf{v^c} \\
		\mathbf{v^p}
	\end{bmatrix}.
\end{equation}

Using the above definitions, the imaging model can be expressed as
\begin{equation}
	\label{eq37}
	\mathbf{y} = \boldsymbol{\Phi} \mathbf{x} + \mathbf{v},
\end{equation}
where \(\boldsymbol{\Phi}\) is a sensing matrix in \(\mathbb{R}^{M \times N}\) that encodes the system mapping from the original scene \(\mathbf{x}\) to the measurements \(\mathbf{y}\). \(\mathbf{y} \in  \mathbb{R}^{M \times 1}\), \(\mathbf{x} \in  \mathbb{R}^{N \times 1}\), and \(\boldsymbol{\Phi} \in  \mathbb{R}^{M \times N}\), The dimensions \(M\) and \(N\) are defined as follows: \(M\) is calculated as \(H(W+d(C-1))+HW\) and \(N\) is calculated as \(HWC\), where \(d\) represents the dispersion step size. It is important to note that in the context of HSI reconstruction, \(M\) is typically much smaller than \(N\), highlighting that the HSI reconstruction task in DCCHI is an ill-posed problem.

\begin{figure*}[!t]
	\centering
	\includegraphics[width=5.8in,height=2.4in]{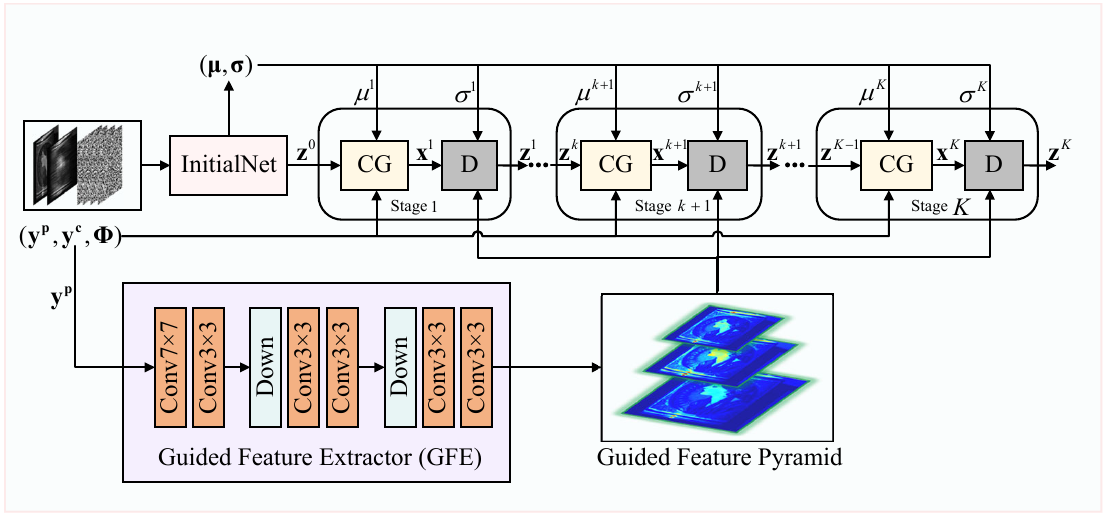} 
	\caption{\textbf{Overview of PGDU.} The InitialNet initiates the process with compressive measurements and sensing matrix, followed by a series of stages each containing a conjugate gradient (CG) block and an In2SET denoiser.} 
	\label{fig:framework}
    \vspace{-2.5mm}
\end{figure*}

\subsection{Reconstruction Framework}
Compared to learn a brute-force mapping between HSI $\mathbf{x}$ and measurement $\mathbf{y}$, model-based methods have potential in compressive imaging due to fully considering the imaging model. Model-based methods usually fomulate HSI reconstruction as a Bayesian inference challenge, solving Eq.~\eqref{eq37} under a unified Maximum A Posteriori (MAP) framework. Mathematically, the optimization problem for HSI reconstruction could be expressed as
\vspace{-0.8mm}
\begin{equation}
\vspace{-0.8mm}
	\label{eq41}
	\hat{\mathbf{x}}=\arg \min _{\mathbf{x}} \frac{1}{2}\|\mathbf{y}-\boldsymbol{\Phi} \mathbf{x}\|_2^2+\eta J(\mathbf{x}),
\end{equation}
where $J(\mathbf{x})$ represents regularization. The parameter $\eta$ is a regularization coefficient, controlling the trade-off between data fidelity and the regularization term imposed by $J(\mathbf{x})$.

To solve the optimization problem shown in Eq.~\eqref{eq41}, we design the PGDU based on the physical structure of DCCHI, as shown in Figure~\ref{fig:framework}. Utilizing the half quadratic splitting (HQS) method, we decompose the optimization problem into two sub-problems: the data fidelity term, as shown in Eq.~\eqref{eq4_data}; and the prior term, as shown in Eq.~\eqref{eq4_prior}.
\vspace{-0.8mm}
\begin{equation}
	\label{eq4_data}
	\mathbf{x}^{k+1} = \arg \min _{\mathbf{x}}\left\|\mathbf{y}-\boldsymbol{\Phi} \mathbf{x}\right\|_2^2 + \mu^{k+1}\left\|\mathbf{z}^{k}-\mathbf{x}\right\|_2^2,
\end{equation}
\vspace{-1.0mm}
\begin{equation}
	\label{eq4_prior}
	\mathbf{z}^{k+1} = \arg \min _{\mathbf{z}} \frac{1}{2} \mu^{k+1}\left\|\mathbf{z}-\mathbf{x}^{k+1}\right\|_2^2 + \eta^{k+1} J\left(\mathbf{z}\right),
\end{equation}
the superscript \(k\) indicates the stage index number, and \(\mathbf{z}\) is an auxiliary variable introduced to facilitate the optimization. The parameter \(\mu^{k+1}\)  is a relaxation parameter that governs the alignment between the current estimate \(\mathbf{x}\) and the auxiliary variable \(\mathbf{z}\). The parameter \(\eta^{k+1}\) is a stage-specific parameter for the regularization coefficient \(\eta\).

\begin{figure*}[!t]
	\centering
	\includegraphics[width=6.8in,height=3.2in]{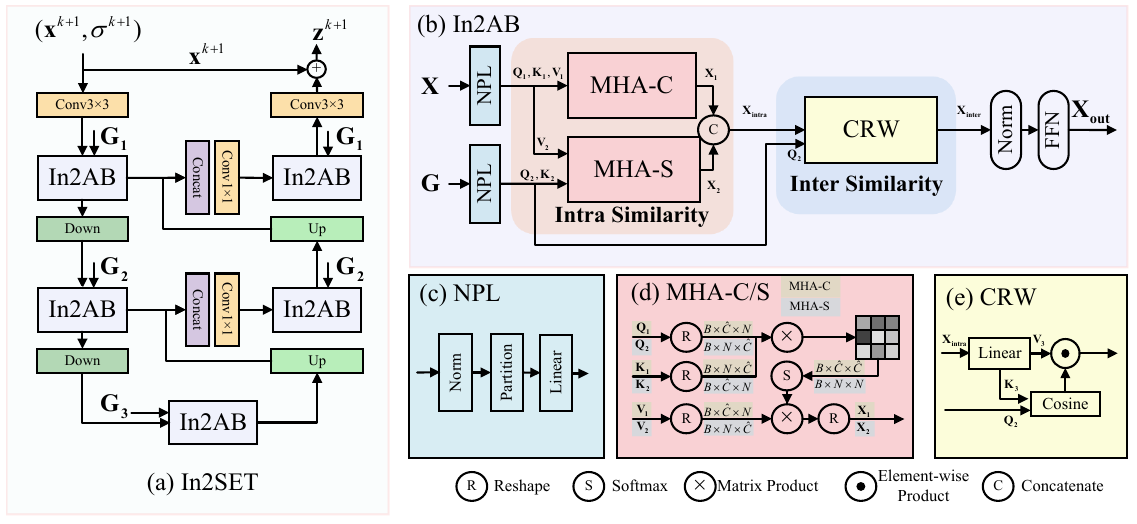} 
	\caption{Diagram of In2SET architecture. (a) The U-shaped In2SET structure. (b) In2AB, consisting of two normalization layers, an intra-similarity attention module, an inter-similarity attention module, and an FFN layer. (c) The components of NPL. (d) The multi-head self-attention in channel (MHA-C) and multi-head cross-attention in spatial (MHA-S). (e) The cosine similarity reweighting (CRW) mechanism.} 
	\label{fig:model}
\end{figure*}

The data fidelity term can be directly calculated in closed form, as demonstrated in Eq.~\eqref{eq48}.
\begin{equation}
	\label{eq48}
	\mathbf{x}^{k+1} = \left(\boldsymbol{\Phi}^\mathsf{T} \boldsymbol{\Phi} + \mu \mathbf{I}\right)^{-1}\left(\boldsymbol{\Phi}^\mathsf{T} \mathbf{y} + \mu^{k+1} \mathbf{z}^k\right),
\end{equation}
in the context of HSI reconstruction in CASSI, the matrix \(\boldsymbol{\Phi}^\mathsf{T} \boldsymbol{\Phi}\) is strictly diagonal. However, with the addition of the PAN camera branch in DCCHI, the matrix $\boldsymbol{\Phi}^\mathsf{T} \boldsymbol{\Phi}$ becomes only diagonally dominant, and thus a closed-form solution cannot be explicitly calculated.

Given the non-diagonal form of $\boldsymbol{\Phi}^\mathsf{T} \boldsymbol{\Phi}$, we employ the conjugate gradient (CG) method~\cite{na2009co} for the efficient resolution of the data fidelity term. The CG method is well suited for this problem due to its effectiveness in solving large-scale, sparse linear systems.

To fully explore the guidance of the PAN image $\mathbf{y^p}$ on prior term, we design a PAN-guided Gaussian denoiser to estimate $\mathbf{z}^{k+1}$. This is formulated as Eq.~\eqref{eq45}:
\begin{equation}
	\label{eq45}
	\mathbf{z}^{k+1} = \mathrm{Denoiser}_{k+1}\left(\mathbf{x}^{k+1}, \mathbf{GFE(y^p)},\sigma^{k+1}\right),
\end{equation}
where $\mathbf{GFE(y^p)}$ serves as structural guidance for the denoising, which incorporates multi-scale spatial details from the high-resolution $\mathbf{y^p}$ data into the denoiser, aiming to refine the fidelity of the reconstructed HSI by adding contextual information. The Gaussian noise level of the denoiser, denoted as $\sigma^{k+1}$, is equal to $\sqrt{\eta^{k+1} / \mu^{k+1}}$. The $\mu^{k+1}$ and $\sigma^{k+1}$ are the elements of vectors $\boldsymbol{\mathbf{\mu}}$ and $\boldsymbol{\mathbf{\sigma}}$, which are inferred by the InitialNet of PGDU.

The feature extraction from $\mathbf{y^p}$ is delineated as follows: within the unrolling framework, the feature representation of $\mathbf{y^p}$ is extracted singularly and iteratively shared across multiple stages. A stack of convolutional blocks is utilized to harvest feature maps at varying resolutions, assembling a guided feature pyramid $[\mathbf{G_1}, \mathbf{G_2}, \mathbf{G_3}]$. The guided feature pyramid is integral to the subsequent denoising stage, aiding in the preservation of intricate spatial details during the denoising.

\section{In2SET}
\subsection{Overall Architecture}
In this section, we introduce the denoiser In2SET within the PGDU framework. As shown in Figure~\ref{fig:model} (a), the proposed In2SET adopts a U-shaped architecture~\cite{ro15un}, involving the intra-inter attention block (In2AB) as the foundational units. This architecture employs multi-scale feature extraction coupled with skip connections between the encoder and decoder, enhancing the efficiency in information processing and feature extraction. To be specific, the denoiser initializes feature map $\mathbf{X}$ from the spectral image $\mathbf{x}^{k+1}$ and the noise level $\sigma^{k+1}$, passing through a convolutional layer with a kernel size of $3 \times 3$. The input fed into each In2AB consists of guided feature $\mathbf{G_{i}}$, feature maps extracted from the PAN image, and HSI feature $\mathbf{X}$. The denoised image $\mathbf{z}^{k+1}$ is the sum of $\mathbf{x}^{k+1}$ and the output of the last In2AB, calculated by a convolutional layer with a kernel size of $3 \times 3$.
The In2AB structure is shown in Figure~\ref{fig:model} (b). Our In2AB module includes two normalization layers, one intra-similarity attention module, one inter-similarity attention module, and one feed-forward fully connected (FFN) layer. In the following, we introduce the detail of the intra-similarity and inter-similarity.
\subsection{Intra-Similarity Attention}
\label{sec:submethod}
Our intra-similarity attention module is grounded on two key principles: firstly, for exploring intra-spatial similarity, the comprehensive and continuous spatial data provided by the PAN image are invaluable. This allows the spatial similarity in the PAN image to serve as an effective proxy for approximating the spatial similarity of the HSI. Secondly, in addressing intra-spectral similarity, analyzing the reconstructed HSI becomes significantly more effective, given the PAN image's deficiency in spectral color information.

As shown in Figure ~\ref{fig:model} (b), our intra-similarity attention module includes two branches: multi-head self-attention in channel (MHA-C) and multi-head cross-attention in spatial (MHA-S), representing the exploration of spectral and spatial self-similarity, respectively. The MHA-C is based on the work in MST~\cite{ca22ma}, which has been demonstrated to effectively explore spectral self-similarity.

\begin{table*}[t]
	\centering
	{\fontsize{7.8pt}{9.0pt}\selectfont 
		\setlength{\tabcolsep}{4.50pt}
		\begin{tabular}{@{}p{3.0cm}cccccccccccc@{}}
			\hline
			\multirow{2}{*}{Method} & \multirow{2}{*}{GFLOPs} & \multicolumn{10}{c}{Scene} & \multirow{2}{*}{Average}\\
			\cline{3-12}
			& & 01 & 02 & 03 & 04 & 05 & 06 & 07 & 08 & 09 & 10 & \\
			\hline
			\multirow{2}{*}{PFusion-RGB~\cite{he21fa}} & \multirow{2}{*}{-} & 40.09 & 38.84 & 38.70 & 46.65 & 32.07 & 37.12 & 39.74 & 36.75 & 34.52 & 35.53 & 38.00\\
			&  & 0.979 & 0.968 & 0.966 & 0.936 & 0.980 & 0.980 & 0.964 & 0.965 & 0.931 & 0.979 & 0.965\\
			\hline
			\multirow{2}{*}{PIDS-RGB~\cite{ch23pr}} & \multirow{2}{*}{-} & 42.09 & 40.08 & 41.50 & 48.55 & 40.05 & 39.00 & 36.63 & 37.02 & 38.82 & 38.64 & 40.24\\
			&  & 0.983 & 0.949 & 0.968 & 0.989 & 0.982 & 0.974 & 0.940 & 0.948 & 0.953 & 0.980 & 0.967\\
			\hline
			\multirow{2}{*}{TV-DC~\cite{wa15du}} & \multirow{2}{*}{-} & 35.81 & 33.22 & 31.07 & 40.11 & 33.32 & 34.62 & 31.09 & 32.31 & 29.36 & 33.84 & 33.47\\
			&  & 0.947 & 0.885 & 0.879 & 0.947 & 0.944 & 0.943 & 0.885 & 0.916 & 0.862 & 0.953 & 0.910\\
			\hline       
			\multirow{2}{*}{PIDS-DC~\cite{ch23pr}} & \multirow{2}{*}{-} & 39.82 & 37.07 & 37.72 & 46.78 & 37.45 & 37.74 & 32.90 & 31.66 & 34.35 & 38.58 & 37.41\\
			&  & 0.977 & 0.921 & 0.950 & 0.978 & 0.973 & 0.963 & 0.896 & 0.915 & 0.902 & 0.972 & 0.945\\
			\hline
			\multirow{2}[2]{*}{BiSRNet-DC~\cite{ca23bi}} & \multirow{2}[2]{*}{1.33 } & 35.02  & 34.13  & 31.50  & 35.88  & 33.70  & 35.58  & 32.31  & 32.73  & 31.37  & 34.48  & 33.67  \\
			&       & 0.945  & 0.914  & 0.883  & 0.895  & 0.935  & 0.925  & 0.900  & 0.903  & 0.899  & 0.936  & 0.914  \\
			\hline
			\multirow{2}[2]{*}{CST-DC~\cite{ca22co}} & \multirow{2}[2]{*}{25.40 } & 37.44  & 38.91  & 36.79  & 42.27  & 36.57  & 38.91  & 36.87  & 35.91  & 35.87  & 37.93  & 37.75  \\
			&       & 0.975  & 0.978  & 0.969  & 0.983  & 0.982  & 0.984  & 0.967  & 0.980  & 0.973  & 0.990  & 0.978  \\
			\hline
			\multirow{2}[2]{*}{HDNet-DC~\cite{hu22hd}} & \multirow{2}[2]{*}{144.31 } & 38.06  & 39.79  & 38.21  & 42.79  & 37.22  & 39.26  & 37.41  & 36.51  & 36.64  & 37.52  & 38.34  \\
			&       & 0.976  & 0.982  & 0.973  & 0.983  & 0.983  & 0.986  & 0.968  & 0.982  & 0.976  & 0.988  & 0.980  \\
			\hline
			\multirow{2}[2]{*}{MST-DC~\cite{ca22ma}} & \multirow{2}[2]{*}{25.77 } & 38.38  & 40.45  & 37.63  & 42.88  & 37.72  & 39.66  & 37.56  & 37.40  & 37.86  & 38.10  & 38.76  \\
			&       & 0.976  & 0.976  & 0.966  & 0.975  & 0.978  & 0.976  & 0.965  & 0.971  & 0.971  & 0.981  & 0.974  \\
			\hline
			\multirow{2}[2]{*}{MST++-DC~\cite{ca22mst++}} & \multirow{2}[2]{*}{17.69 } & 38.38  & 40.47  & 37.70  & 43.88  & 37.75  & 39.42  & 37.48  & 37.38  & 38.82  & 39.04  & 39.03  \\
			&       & 0.977  & 0.980  & 0.968  & 0.984  & 0.983  & 0.985  & 0.964  & 0.982  & 0.980  & 0.989  & 0.980  \\
			\hline
			\multirow{2}{*}{DAUHST-DC-2stg~\cite{ca22de}} & \multirow{2}{*}{16.79} & 40.78 & 43.13 & 41.73 & 47.09 & 39.84 & 40.90 & 39.75 & 38.98 & 41.29 & 40.04 & 40.22\\
			&  & 0.983 & 0.987 & 0.980 & 0.990 & 0.987 & 0.986 & 0.976 & 0.981 & 0.983 & 0.988 & 0.983\\
			\hline
			\multirow{2}{*}{DAUHST-DC-3stg~\cite{ca22de}} & \multirow{2}{*}{24.70} & 40.22 & 43.52 & 41.74 & 47.07 & 38.81 & 40.16 & 39.86 & 38.21 & 40.63 & 39.32 & 40.95\\
			&  & 0.983 & 0.989 & 0.981 & 0.993 & 0.985 & 0.987 & 0.978 & 0.981 & 0.983 & 0.990 & 0.985\\
			\hline
			\multirow{2}{*}{DAUHST-DC-5stg~\cite{ca22de}} & \multirow{2}{*}{40.51} & 40.74 & 44.00 & 41.58 & 46.84 & 39.66 & 40.89 & 40.21 & 38.72 & 39.98 & 40.10 & 41.27\\
			&  & 0.984 & 0.989 & 0.981 & 0.991 & 0.986 & 0.987 & 0.979 & 0.983 & 0.982 & 0.989 & 0.916\\
			\hline
			\multirow{2}{*}{DAUHST-DC-9stg~\cite{ca22de}} & \multirow{2}{*}{72.11} & 41.59 & 45.19 & 43.47 & 48.92 & 40.27 & 41.17 & 40.73 & 40.11 & 43.50 & 41.33 & 42.62\\
			&  & 0.985 & 0.991 & 0.984 & 0.993 & 0.988 & 0.988 & 0.979 & 0.986 & 0.988 & 0.990 & 0.987\\
			\hline
			\rowcolor{yellow!20} &  & 40.33 & 42.30 & 40.34 & 47.24 & 39.42 & 40.61 & 39.46 & 38.42 & 40.37 & 39.96 & 40.84\\
			\rowcolor{yellow!20} \multirow{-2}{*}{In2SET-2stg (Ours)} & \multirow{-2}{*}{14.35} & 0.983 & 0.985 & 0.977 & 0.991 & 0.986 & 0.986 & 0.975 & 0.979 & 0.981 & 0.988 & 0.983\\
			\hline
			\rowcolor{yellow!20} &  & 40.78 & 43.13 & 41.73 & 47.09 & 39.84 & 40.90 & 39.75 & 38.98 & 41.29 & 40.04 & 41.35\\
			\rowcolor{yellow!20} \multirow{-2}{*}{In2SET-3stg (Ours)} & \multirow{-2}{*}{20.79} & 0.983 & 0.987 & 0.980 & 0.990 & 0.987 & 0.986 & 0.976 & 0.981 & 0.983 & 0.988 & 0.984\\
			\hline
			\rowcolor{yellow!20} & & 41.13 & 44.43 & 42.74 & 47.29 & 40.33 & 40.95 & 40.49 & 39.15 & 42.07 & 39.44 & 41.80\\
			\rowcolor{yellow!20} \multirow{-2}{*}{In2SET-5stg (Ours)} & \multirow{-2}{*}{33.66} & 0.985 & 0.990 & 0.983 & 0.993 & 0.988 & 0.987 & 0.979 & 0.982 & 0.985 & 0.987 & 0.985\\
			\hline
			\rowcolor{yellow!20} & & \textbf{42.56} & \textbf{46.42} & \textbf{44.55} & \textbf{50.63} & \textbf{42.01} & \textbf{42.49} & \textbf{41.59} & \textbf{40.53} & \textbf{43.83} & \textbf{42.33} & \textbf{43.69}\\
			\rowcolor{yellow!20} \multirow{-2}{*}{In2SET-9stg (Ours)} & \multirow{-2}{*}{59.40} & \textbf{0.989} & \textbf{0.994} & \textbf{0.986} & \textbf{0.996} & \textbf{0.992} & \textbf{0.991} & \textbf{0.983} & \textbf{0.989} & \textbf{0.990} & \textbf{0.994} & \textbf{0.990}\\
			\hline
	\end{tabular}}
	\caption{Comparison of In2SET with SOTA DCCHI methods across 10 simulated scenes, including FLOPS, PSNR (upper entry), and SSIM (lower entry). ``Method-RGB'' indicates the use of RGB observations, while ``Method-DC'' denotes grayscale observations in PAN images. It is noted that ``Method-DC'' is modified from other HSI reconstruction method for the DCCHI reconstruction task.}
	\label{table:quantitative_results}
    \vspace{-2.5mm}
\end{table*}

The input features $\mathbf{X}$ and $\mathbf{G}$ have dimensions $\mathbb{R}^{H \times W \times C}$ and $\mathbb{R}^{H \times W \times \hat{C}}$, respectively, where $\hat{C}=\frac{C}{2}$. After normalization and partitioning, the dimensions become \( \mathbb{R}^{B \times N \times C} \) and \( \mathbb{R}^{B \times N \times \hat{C}} \). In the first and last layers of HSAB within In2SET, \( B = \frac{HW}{M^2}\)  and \( N =M^2 \) . For the remaining layers, dimensions are recalibrated to \( B = M^2 \) and \( N = \frac{HW}{M^2} \). Following linear projection as shown in Eq.~(\ref{eqL1}) and Eq.~(\ref{eqL2}).
\begin{equation}
	\label{eqL1}
	\mathbf{Q_1},\mathbf{K_1},\mathbf{V_1},\mathbf{V_2} = L_{Q_1,K_1,V_1,V_2}(\mathbf{X}),
\end{equation}
\begin{equation}
	\label{eqL2}
	\mathbf{Q_2},\mathbf{K_2} = L_{Q_2,K_2}(\mathbf{G}),
\end{equation}
where $\mathbf{X}$ is projected to $\mathbf{Q_1},\mathbf{K_1},\mathbf{V_1} \in  \mathbb{R}^{B \times \hat{C} \times N} $ and $\mathbf{V_2} \in  \mathbb{R}^{B \times N \times \hat{C}} $. $\mathbf{G}$ is projected to $\mathbf{Q_2},\mathbf{K_2} \in  \mathbb{R}^{B \times N \times \hat{C}} $.

The computations of MHA-C and MHA-S can be represented by Eq.~(\ref{eqA1}) and Eq.~(\ref{eqA2}),
\begin{equation}
	\label{eqA1}
	\mathbf{X_1} = \operatorname{Softmax}\left(\frac{\mathbf{Q}_1\mathbf{K}_1^T}{\sqrt{d_{h1}}}+\mathbf{P}_1\right) \mathbf{V}_1,
\end{equation}
\begin{equation}
	\label{eqA2}
	\mathbf{X_2} = \operatorname{Softmax}\left(\frac{\mathbf{Q}_2\mathbf{K}_2^T}{\sqrt{d_{h2}}}+\mathbf{P}_2\right) \mathbf{V}_2,
\end{equation}
where $\sqrt{d_{h1}}$ and $\sqrt{d_{h2}}$ are scalars as defined in~\cite{ca22de}, $\mathbf{P}_1 \in  \mathbb{R}^{B \times \hat{C} \times \hat{C}}$, and $\mathbf{P}_2 \in  \mathbb{R}^{B \times N \times N}$ denote learnable positional encodings. 

The output of intra-attention $\mathbf{X_{intra}}$ can be obtained by Eq.~(\ref{eqIntra}).
\begin{equation}
	\label{eqIntra}
	\mathbf{X_{intra}} = \operatorname{Concat}\left(\mathbf{X_1},\mathbf{X_2}\right).
\end{equation}

\subsection{Inter-Similarity Attention}
We establish our inter-similarity attention module based on one principle: the shallow semantic features of the PAN image and the corresponding HSI area are similar, including consistency in texture, shape features, and low-level image attributes (brightness and contrast). The credibility of the reconstructed area is directly proportional to this consistency. Inter-similarity attention explores cross-modal similarity using cosine similarity reweighting (CRW).

After exploiting intra-similarity, the output of the intra-similarity attention module $\mathbf{X_{intra}}$ is projected to obtain $\mathbf{K_3}$ and $\mathbf{V_3}$,
\begin{equation}
	\label{eq:L3}
	\mathbf{K_3}, \mathbf{V_3} = L_{K_3,V_3}(\mathbf{X_{intra}}),
\end{equation}
where $\mathbf{K_3} \in \mathbb{R}^{B \times N \times \hat{C}}$ and $\mathbf{V_3} \in \mathbb{R}^{B \times N \times C}$.

Then, the computations of CRW can be represented as follows:
\begin{equation}
	\label{eq:Inter}
	\mathbf{X_{inter}} = \frac{\mathbf{Q_2} \cdot \mathbf{K_3}}{\| \mathbf{Q_2} \| \| \mathbf{K_3} \|} \odot \mathbf{V_3},
\end{equation}
where the cosine similarity computed along the last dimension between $\mathbf{Q_2}$ and $\mathbf{K_3}$ quantifies the feature alignment in corresponding areas. A higher similarity score reflects a stronger feature correspondence, which, in turn, informs the enhanced weighting of $\mathbf{V_3}$, improving the fidelity of the region being reconstructed. 

\section{Experiments}
In this section, we begin by introducing detailing the network training process. Subsequently, we evaluate the performance of In2SET on both simulation and real-world DCCHI datasets. Then, we validate the effectiveness of the intra-similarity block using statistical simulation data. Finally, we perform an ablation study to demonstrate the effectiveness of our proposed method.

\textbf{Simulation Dataset.} Two distinct datasets are used: CAVE~\cite{ya10ge} and KAIST~\cite{ch17hi}. The CAVE dataset is made up of 32 hyperspectral images with \(512 \times 512\) spatial resolution. The KAIST dataset contains 30 hyperspectral images, each with a larger spatial dimension of \(2704 \times 3376\). The CAVE dataset is used as the training data, while a subset of 10 scenes crop from the KAIST dataset is used for the evaluation phase, in accordance with the protocols established in references~\cite{me20en,hu21de,ca22ma,hu22hd,ca22de}.

\textbf{Implementation}. Our method is implemented with Pytorch and trained with Adam~\cite{ki14ad} optimizer for 300 epochs. During training, the learning rate is $4\times10^{-4}$ using the cosine annealing scheduler, and the loss function for network training is $L_1$ loss. We randomly extract patches from 3D HSI cubes to serve as training samples. The dimensions of these patches are $256 \times 256 \times 28$ for the simulation experiment and $350 \times 260 \times 26$ for the real-world experiment. In simulation imaging model, we configure a dispersion shift step $d$ of 2, directing the dispersion to the right. In real-word imaging model, the dispersion shift step d is set to 1, with the dispersion oriented upwards.
\vspace{-1.00mm}
\subsection{Simulation Data Results}
\vspace{-1.00mm}
The reconstruction quality of hyperspectral images are assessed using peak signal-to-noise ratio (PSNR) and structural similarity index (SSIM)~\cite{wa04im} metrics. 

The Table~\ref{table:quantitative_results} clearly shows that In2SET excels in all test scenarios, particularly in In2SET-9stg, achieving the best performance across all scenes with an average PSNR of 43.69 dB and a SSIM of 0.990. This significantly surpasses model optimization methods like TV-DC and PIDS-DC. Compared to deep learning-based methods, such as various stages of DAUHST-DC~\cite{ca22de}, In2SET demonstrates remarkable superiority. In2SET-3stg achieves performance comparable to that of DAUHST-DC-5stg~\cite{ca22de}, but with only 51.3\% of the cost, amounting to 20.78 GFLOPs compared to 40.50 GFLOPs.

\begin{figure*}[t]
	\centering
	\includegraphics[width=6.8in,height=2.5in]{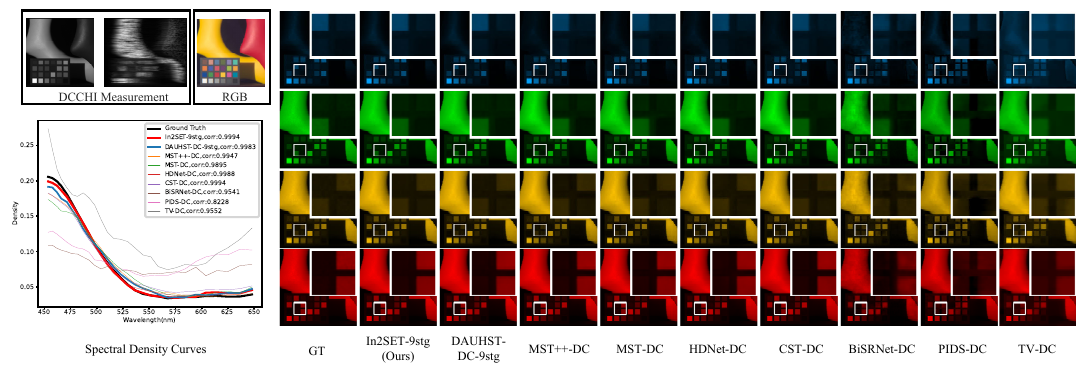} 
	\caption{Comparative reconstruction results of different reconstruction methods for Scene 3 from the KAIST dataset at spectral bands 476.5nm, 536.5nm, 584.5nm, and 625.0nm. The spectral density curves are plotted from the blue region in the colorchecker.} 
	\label{fig:residual_heatmaps}
	\vspace{-5mm} 
\end{figure*}

To facilitate a direct visual quality assessment, we analyzed the distinction between Ground Truth (GT) and the reconstructions produced by various open-source methods. Figure~\ref{fig:residual_heatmaps} presents reconstruction results for selected spectral bands. These results are generated from two model-based approaches, TV-DC~\cite{wa15du} and PIDS-DC~\cite{ch23pr}, five end-to-end network algorithms, BiSRNet-DC~\cite{ca23bi}, CST-DC~\cite{ca22co}, HDNet-DC~\cite{hu22hd}, MST-DC~\cite{ca22ma}, MST++-DC~\cite{ca22mst++}, as well as from two deep unfolding methods, namely DAUHST-DC-9stg~\cite{ca22de} and our In2SET-9stg. The comparison underlines the exceptional ability of our In2SET approach to maintain spatial and spectral fidelity. 

We suggest that the intra similarity in PAN image can effectively approximates that in HSI. To support this, we experiment with three public datasets: CAVE~\cite{ya10ge}, KAIST~\cite{ch17hi} and ICVL~\cite{ar16sp}. We compute spatial correlation maps for each scene, comparing the correlation map of the HSI with that of the corresponding PAN image. The results of this comparison are displayed in Table~\ref{table:corr}, indicating a high degree of similarity between the intra-spatial correlations in PAN images and HSI. More visual results are presented in the supplementary materials.
\vspace{-1.00mm}
\subsection{Real Data Results}
\begin{figure}[!t]
	\centering
	\includegraphics[width=2.908in,height=3.5in]{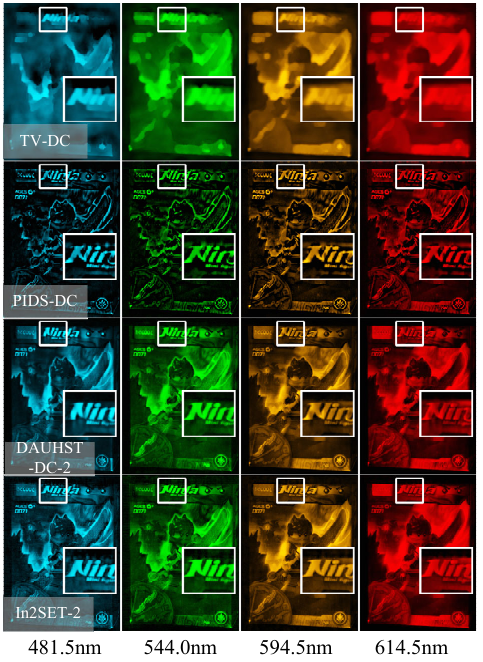} 
	\caption{Visualization of reconstruction performance across different spectral bands on real data. The spectral bands selected for comparison are 481.5nm, 544.0nm, 594.5nm, and 614.5nm. 
	} 
	\label{fig:ninja_reconstruction_comparison}
	\vspace{-6mm} 
\end{figure}

\vspace{-1.00mm}
In this research, we used a real-world DCCHI measurement \textit{Ninja}, taken from publicly available data as detailed in reference~\cite{wa17ad}. Figure~\ref{fig:ninja_reconstruction_comparison} illustrates the reconstruction results for four spectral bands in this scene, using various DCCHI reconstruction algorithms. The comparison highlights the superior image restoration quality of our model over other methods, validating its effectiveness and reliability in real-world applications.

\vspace{-1.30mm}
\subsection{Ablation Study}
\vspace{-1.00mm}
To verify the effectiveness of our proposed method, we conducted ablation studies for the In2SET method. All evaluations are conducted on simulated datasets.
\begin{table}[t]
  \centering
  \small
  \setlength{\tabcolsep}{13.5pt}
  \setstretch{1.00}
\begin{tabular}{llll}
	\hline
	        & RMSE  & Correlation & PSNR(dB)  \\ \hline
	CAVE    & 0.034 & 0.999       & 36.28 \\     \hline
	KAIST   & 0.025 & 0.996       & 35.74 \\     \hline
	ICVL    & 0.076 & 0.997       & 31.10 \\
	\hline
\end{tabular}
\vspace{-3mm}
\caption{Comparison of correlation maps between HSI and PAN image, including average Root Mean Squared Error (RMSE), average correlation, and PSNR.} 
\label{table:corr}
\vspace{-3mm} 
\end{table}
\begin{table}[t]
  \centering
  \small
  \setlength{\tabcolsep}{3.0pt}
  \setstretch{1.0}
  \begin{tabular}{@{}ccccccp{1.0cm}<{\centering}p{1.0cm}<{\centering}@{}}
    \hline
    Baseline & CRW & MHA-C & MHA-S & PSNR(dB) & SSIM & GFLOPs\\
    \hline
    \checkmark &  &  &  & 37.23 & 0.944 & 9.87\\
    \hline
    \checkmark & \checkmark &  &  & 40.05 & 0.980 & 11.19\\
    \hline
    \checkmark & \checkmark & \checkmark &  & 40.46 & 0.982 & 14.06\\
    \hline
    \checkmark & \checkmark & \checkmark & \checkmark & 40.84 & 0.983 & 14.35\\
    \hline
  \end{tabular}
  \vspace{-3mm}
  \caption{Break-down ablation study on individual components of the proposed method.}
  \label{table:ablation1}
  \vspace{-3mm} 
\end{table}
\begin{table}[t]
  \centering
  \small
  \setlength{\tabcolsep}{13.5pt}
  \setstretch{1.0}
  \begin{tabular}{@{}ccccc@{}}
    \hline
    Methods & CG-1 & CG-2 & CG-5 & CG-10\\
    \hline
    PSNR(dB) & 40.42 & 40.66 & 40.84 & 41.05\\
    \hline
    SSIM & 0.982 & 0.983 & 0.983 & 0.984\\
    \hline
    FPS & 26.45 & 22.79 & 19.16 & 13.99\\
    \hline
  \end{tabular}
  \vspace{-3mm}
  \caption{Ablation study on iterative performance of conjugate gradient descent for data item.}
  \label{table:ablation2}
  \vspace{-5mm} 
\end{table}

\textbf{Break-down Ablation.} The break-down study, as delineated in Table~\ref{table:ablation1}, offers a detailed examination of how each component within our proposed method influences reconstruction performance. Starting from a baseline derived by removing all ablated components from In2SET-2stg, the model achieves a PSNR of 37.23 dB. The addition of the CRW component leads to a significant increase in PSNR by +2.82 dB, albeit at an increased computational cost of +1.32 GFLOPs. The subsequent integration of MHA-C further elevates the PSNR by +0.41 dB. The final inclusion of MHA-S brings an additional PSNR gain of +0.38 dB. These step-by-step enhancements highlight the critical role each component plays in enhancing the overall capability of the model for HSI reconstruction, demonstrating a progressive improvement in performance with each added component.

\textbf{Ablation Study of CG Iterations.} Table~\ref{table:ablation2} presents a comparative analysis of PSNR, SSIM, and Frames Per Second (FPS) metrics across varying numbers of CG iterations for HSI reconstruction. The CG-1, which conducts a single iteration, essentially operates as a vanilla gradient descent. These experiments are conducted on a system equipped with a TITAN Xp GPU (12GB) and an Intel (R) Xeon (R) Platinum 8358P CPU @ 2.60GHz. 

There is a clear trend of improvement in PSNR as the number of CG iterations increases from 1 to 10, with values escalating from 40.42 dB to 41.05 dB. However, an increase in iterations from CG-1 to CG-10 leads to a decrease in FPS from 26.45 dB to 13.99 dB, indicating a trade-off between reconstruction quality and computational speed. A notable observation is that CG-5 offers an optimal balance between time efficiency and performance, making it the preferred choice for our In2SET network. 
\vspace{-1.0mm}
\section{Conclusion}
\vspace{-1.0mm}
In this paper, we propose the In2SET for DCCHI reconstruction. Our method maximizes PAN image utility for HSI reconstruction by: (1) approximating HSI intra-similarity using PAN image, addressing the unreliability of direct computation from intermediate results. (2) Leveraging inter-similarity between HSI and PAN images to accurately reconstruct regions, providing cues for uncertain areas. Integrating In2SET into a PGDU framework allowed us to substantially enhance the spatial-spectral fidelity and detail of reconstructed images. Experiments on real and simulated datasets show that our method consistently outperforms the state-of-the-art while maintaining lower computational complexity.

\noindent\textbf{Acknowledgments} This work is partially supported by the National Natural Science Foundation of China under grants 62322204, 62131003, 62072038 and 62302041.
{
	\small
	\bibliographystyle{ieeenat_fullname}
	\bibliography{main}
}


\end{document}